\documentclass{svjour3}                     
\usepackage{amssymb}
\usepackage{graphicx,epsfig}
\usepackage{natbib}
\usepackage{mathptmx}      
\bibpunct{(}{)}{;}{a}{}{,}
\makeatletter
\renewcommand*{\@biblabel}[1]{\hfill}
\makeatother

\newcommand{\ion}[2]{#1\,{\sc{#2}}}

\newcommand{\aeta}[3]{  #1, { A\&A}, {  #2}, #3}

\newcommand{\aspj}[3]{  #1, { ApJ}, {  #2}, #3}
\newcommand{\mnras}[3]{   #1, { MNRAS}, {  #2}, #3}
\newcommand{\pasj}[3]{   #1, { PASJ}, {  #2}, #3}

 \journalname{SSRv}

\begin{document}

\title{Soft X-ray and extreme ultraviolet excess emission from
clusters of galaxies }

\author{F.~Durret \and
        J.S.~Kaastra \and
        J.~Nevalainen \and
        T.~Ohashi \and
        N.~Werner}

\authorrunning{Durret et al.}
\titlerunning{Soft excess emission from clusters}

\institute{F. Durret \at Institut d'Astrophysique de Paris, CNRS, UMR~7095,
Universit\'e Pierre et Marie Curie,
98bis Bd Arago, F-75014 Paris, France 
                  \email{durret@iap.fr}\\
J.S. Kaastra \at SRON Netherlands Institute for Space Research, 
Sorbonnelaan 2, 3584 CA Utrecht, the Netherlands \\
Astronomical Institute, Utrecht University, P.O. Box 80000, 
3508 TA Utrecht, The Netherlands \\
J. Nevalainen \at Observatory, P.O. Box 14, 00014 University of Helsinki, Finland\\
T. Ohashi \at Department of Physics, Tokyo Metropolitan University, 
                  1-1 Minami-Osawa, Tokyo 192-0397, Japan\\
N. Werner \at SRON Netherlands Institute for Space Research, 
Sorbonnelaan 2, NL - 3584 CA Utrecht, the Netherlands \\ 
Max-Planck-Institute f{\"u}r Astrophysik, Karl-Schwarzschild-Strasse 1, 85749, Garching, Germany
}

\date{Received: 17 September 2007; Accepted: 22 October 2007}

\maketitle

\begin{abstract}
An excess over the extrapolation to the extreme ultraviolet and soft
X-ray ranges of the thermal emission from the hot intracluster medium
has been detected in a number of clusters of galaxies.  We briefly
present each of the satellites (EUVE, ROSAT PSPC and BeppoSAX, and
presently XMM-Newton, Chandra and Suzaku) and their corresponding
instrumental issues, which are responsible for the fact that this soft
excess remains controversial in a number of cases.  We then review the
evidence for this soft X-ray excess and discuss the possible
mechanisms (thermal and non-thermal) which could be responsible for
this emission.

\end{abstract}

\section{Introduction}
\label{Introduction} 

The existence of soft excess emission originating from clusters of
galaxies, defined as emission detected below 1~keV as an excess over
the usual thermal emission from hot intracluster gas (hereafter the
ICM) has been claimed since 1996.  Soft excesses are particularly
important to detect because they may (at least partly) be due to
thermal emission from the Warm-Hot Intergalactic Medium, where as much
as half of the baryons of the Universe could be. They are therefore of
fundamental cosmological importance.

Soft excess emission has been observed (and has also given rise to
controversy) in a number of clusters, mainly raising the following
questions: 1)~Do clusters really show a soft excess? 2)~If so, from
what spatial region(s) of the cluster does the soft excess originate?
3)~Is this excess emission thermal, originating from warm-hot
intergalactic gas (at temperatures of $\sim 10^6$~K), or non-thermal,
in which case several emission mechanisms have been
proposed. Interestingly, some of the non-thermal mechanisms suggested
to account for soft excess emission can also explain the hard X-ray
emission detected in some clusters, for example by RXTE and BeppoSAX (also see
\citealt{petrosian2008} - Chapter 10, this volume; \citealt{rephaeli2008} -
Chapter 5, this volume).

Several instruments have been used to search for soft excess emission:
EUVE, ROSAT and BeppoSAX in the 1990's, and presently XMM-Newton,
Chandra and Suzaku. We will briefly present a history of these
detections, emphasizing the difficulties to extract such weak signal
and the underlying hypotheses, and summarising what is known on each
of the observed clusters. For clarity, results will be presented
separately for the various satellites. Finally we summarise and
discuss the various models proposed to account for Extreme Ultraviolet
(hereafter EUV) emission with their pros and cons.

\section{Instrumental issues}
\label{instrum}

\subsection{The {\sl Extreme Ultraviolet Explorer} (EUVE) satellite}

The EUVE satellite was launched in 1992 and operated till 2001,
covering the $50-250$~eV energy range \citep{bowyer1991}. The 
rectangular shape of the Lex/B ($65-248$~eV) filter
resulted in images where the length of one side
of the image greatly exceeds the other side: $\sim$40$^\prime$ in width
and more than 2$^\circ$ in length (see e.g. Fig. 1 in \citealt{durret2002}).

The spatial scale was 13 pixels/arcmin.  Due to the limited
sensitivity of EUVE, exposure times on clusters were typically between
several tens of ks and 1~Ms.

\subsection{ROSAT PSPC}

The Position Sensitive Proportional Counter PSPC instrument on board the
ROSAT satellite first detected the soft excess component of clusters
of galaxies at X-ray wavelengths.  The PSPC had an effective area of
about 200 cm$^{2}$ at 0.28 keV, enabling soft X-ray studies.  It had a
large field of view (hereafter FOV) of 50$^\prime$ radius, covering the
virial radius in most clusters and enabling in most cases the
estimation of the local background.  The low and stable internal
background of the PSPC enabled reliable X-ray measurements at large
radii where the background is important.  A major problem for soft
excess studies with the PSPC is that it did not cover energies above 2
keV. Thus, it could not be used to determine reliably the hot gas
properties, which had to be measured elsewhere. A further limitation
was the low energy resolution which did not allow detection of
possible emission line blends emanating from the soft component. The
angular resolution of the ROSAT PSPC was $\sim 15^{\prime\prime}$.  Details on
the ROSAT PSPC instrument can be found in \citet{briel1996}.

\subsection{XMM-Newton EPIC}

An important change in the study of the soft excess came with the
XMM-Newton satellite.
The XMM-Newton European Photon Imaging Camera (EPIC) instruments PN
and MOS extend the energy band coverage to 10 keV, thus enabling
simultaneous determination of the hot gas and soft excess
component properties.

The large collecting area of the EPIC telescopes ($\sim$1000 cm$^{2}$
at 0.5 keV for the PN) provides the high statistical quality data
necessary to examine the few 10\% soft excess effect on top of the hot
gas emission.  The spectral resolution of the PN at 0.5 keV is 60 eV
(FWHM), rendering it possible to resolve the emission lines emanating
from the soft excess component.  However, the relatively small FOV (15$^\prime$
 radius) of EPIC prevents the study of the cluster outskirts for
the nearest clusters with single pointings.  The usage of offset
pointings introduce the complex stray light problem which complicates
the analysis of weak signals such as the soft excess.  The more
distant clusters, which would be covered out to the virial radius with
a single pointing, are fainter, which reduces the quality of the
signal.  Thus, the XMM-Newton soft excess analysis is mostly limited
to the central regions of nearby clusters. Details on the
characteristics of XMM-Newton can be found in \citet{turner2001} and
\citet{struder2001}, or in \citet{ehle2006}.

A further complication is the strong and flaring particle-induced
detector background. A local background estimate is vital when
analysing weak signals such as the cluster soft excess. This is
usually not available for nearby clusters since they fill the FOV,
and one has to resort to blank-sky based background estimates.
This introduces uncertainties in the analysis, and further limits
the cluster analysis to central regions where the background is not
important.
Together with the FOV limitations, the background problem limits the
usefulness of XMM-Newton for measuring the soft excess in a large
cluster sample.

The on-going calibration work on the PN and MOS instruments resulted in
changes in the derived soft excess properties for a few clusters
\citep{nevalainen2007}.
Thus, the EPIC results have some degree of systematic uncertainty
based on calibration inaccuracies, and definitive results on the
soft excess properties are not yet available (also see Sect.~4.1.2).

\subsection{Chandra}

Chandra, launched in the same year as XMM-Newton (1999), has a similar
eccentric orbit as the latter satellite and therefore suffers from
comparable enhanced background problems. Its angular resolution is
much higher (0.8$^{\prime\prime}$ Half Energy Width) than that 
of XMM-Newton (14$^{\prime\prime}$). The advantage is that for extended
sources like clusters of galaxies subtraction of contaminating
background point sources can be done more accurately, and more importantly,
blurring effects by the point spread function of the telescope can
usually be ignored. These blurring effects were a serious problem for
the BeppoSAX LECS data (Sect. 3.2.2). However, the effective area at
low energies ($E<0.5$~keV) of Chandra is an order of magnitude smaller
than that of XMM-Newton, and time-dependent contamination that affects
in particular the lower energies is a complicating factor in the
analysis of Chandra data.
 
\subsection{Suzaku XIS}

Suzaku is the fifth Japanese X-ray astronomy satellite, launched in
July 2005 \citep{mitsuda2007}. Unfortunately, the main observing
instrument XRS, consisting of X-ray microcalorimeters used in space
for the first time, did not last until the first space observations,
due to the loss of liquid He \citep{kelley2007}. On the other hand,
the XIS instrument, which employs X-ray CCDs with improved performance
for X-ray spectroscopy, is functioning well \citep{koyama2007}. Regarding soft X-ray spectral studies of diffuse sources, XIS
offers the best capability so far achieved with X-ray satellites.

The XIS system consists of 4 units of mirror and detector
combinations. The X-ray mirrors are identical and have a focal length
of 4.5~m and an effective area of about 500~cm$^2$ at 2 keV \citep{serlemitsos2007}. The angular resolution is about $2^\prime$, limited by the
light-weight design of the thin foil mirror.

The 4 CCDs in the focal plane operate jointly during the
observations. The chips are square with an area of 10~mm\,$\times$\,10~mm and the number of pixels is $10^6$. The field of view
is $17^\prime \times 17^\prime$. One chip is back-illuminated, which
gives a superior soft X-ray sensitivity with somewhat poorer energy
resolution and higher background above 7~keV. The other 3 chips
are of standard front illumination type. The typical energy resolution
is 150~eV FWHM at 5.9~keV at the time of launch. The resolution
degraded significantly with time (200~eV after 1 year), and the XIS team
performed a charge injection operation after October 2006 to maintain
the resolution around 170~eV. In November 2006, one of the 3 FI (front
illumination) chips developed excess noise, and it has been switched
off since then. This leaves a total of 3 chips, one BI (back
illumination) and 2 FIs, in operation. Its background is lower than
that of XMM-Newton.

\section{Soft X-ray excess emission based on EUVE and ROSAT PSPC data}
\label{history}

\subsection{The first objects with a soft X-ray excess discovered: 
Virgo (redshift $z=0.0038$) and Coma ($z=0.0231$)}
\label{virgo_coma}

\begin{figure}    
\begin{center}
\includegraphics[width=\textwidth]{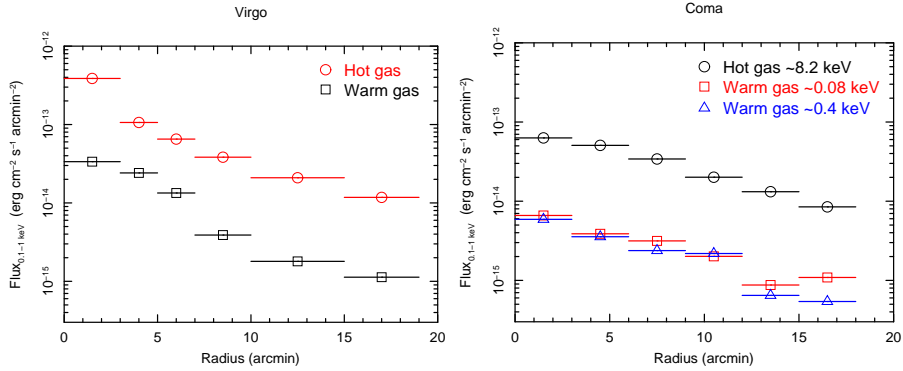}
\caption{ {\emph{Left panel: }} Radial profiles of the $0.1-1.0$~keV
surface brightness of the hot ICM and of the warm gas in Virgo found
by \protect\citet{lieu1996a}. {\emph{Right panel: }} Radial profile of the
$0.1-1.0$~keV surface brightness of the hot ICM and of the two warm gas
components in Coma found by \protect\citet{lieu1996b}. }
\label{fig_Virgo_Coma}
\end{center}  
\end{figure}

The first mention of soft excess emission in the $0.065-0.245$ keV band
was made by \citet{lieu1996a} for the Virgo cluster, based on data
obtained with the EUVE satellite. These authors detected emission
around M~87 up to a radial distance of $\sim$20$^\prime$, and found
excess emission over the best fit single temperature plasma model
obtained by fitting simultaneously the EUVE and ROSAT PSPC
$0.18-2.0$~keV data. This excess was found to decrease with radial
distance (see Fig.~\ref{fig_Virgo_Coma}).  
Similar results were then found for
the Coma cluster, also based on EUVE and ROSAT data, by Lieu et
al. (1996b).

The first interpretation proposed was that the soft excess emission was
thermal, with a single component of ICM gas at a temperature 
between $5\times10^5$~K and $10^6$~K for Virgo \citep{lieu1996a}
and two components at $8\times10^5$~K and $2\times10^6$~K for Coma \citep{lieu1996b}
. The
immediate implication of this hypothesis was that such warm gas would
cool very rapidly, implying a very high mass accretion rate of several
hundred Solar masses per year, and therefore the production of a large
amount of gas in the cool phase (of the order of $10^{14}$~M$_\odot$,
\citealt{lieu1996a,lieu1996b}).

A second model was then proposed by \citet{sarazin1998}, in which
soft excess emission was due to the Inverse Compton emission
(hereafter IC) of cosmic microwave background photons on a relic
population of cosmic ray electrons. Such electrons could have been
produced by supernovae, by radio galaxies or by particle acceleration
in intracluster shocks. Their energy would amount to $1-10$~\% of the
thermal energy content of the ICM. The inverse Compton model was also
favoured by \citet{bowyer1998}, and \citet{lieu1999a}
suggested that cosmic rays could be as energetically important as the
thermal ICM.

From ROSAT images at low energy, \citet{bonamente2001a} derived the
existence of large amounts of cold gas in the ICM of Virgo and Coma,
confirming the existence of a multiphase ICM.

More clusters were then observed, Virgo and Coma were reobserved, and
controversy began to arise concerning the very existence of a soft 
excess over the thermal emission from hot gas responsible for
cluster X-ray emission.

An agreement on the existence of soft excess emission based on EUVE
and ROSAT PSPC data was reached by different observers only on the
Virgo and Coma clusters.  By taking into account the variations of
telescope sensitivity over the field of view, \citet{bowyer1999}
confirmed the presence of EUV excess emission in Coma.  These authors
also later confirmed the soft excess in Virgo and interpreted it as
due to inverse Compton radiation \citep{berghofer2000a}. With EUVE
{\it in situ} background measurements, \citet{bonamente2001b} agreed
with the existence of an EUV excess in Virgo and found that this
excess increased with radius.

Using PSPC data, \citet{bonamente2003} found a very large scale
(2.6~Mpc) soft excess halo in the Coma cluster,
exceeding the thermal emission by 200~\% at the largest radii. Note that
at a distance of 40$^\prime$  the soft excess flux exceeds that of the
hot gas by 100~\%, consistent with the later XMM-Newton analysis of
\citet{finoguenov2003} at the same distance.  The non-thermal model
gave a poor fit to the spectrum of the soft excess, while the thermal
model was acceptable, with a best-fit temperature of $\sim$0.2~keV.

\subsection{Other clusters observed with EUVE and ROSAT}

\subsubsection{Abell~1795 ($z=0.0625$)}
\label{a1795}

A strong soft excess was found in Abell~1795 from EUVE data by \citet{mittaz1998}.  Later, \citet{bonamente2001b} confirmed the
existence of this excess by observing with EUVE an {\it in situ}
background, and found that, as for Virgo, the EUV excess increased
with radius.

However, these results were debated: by taking into account the
variations of telescope sensitivity over the field of view, \citet{bowyer1999} claimed that there was no evidence for EUV excess in
Abell~1795 and later confirmed this result from BeppoSAX data
\citep{berghofer2002}. However, the existence of soft excess
emission in this cluster was confirmed by \citet{durret2002} from
EUVE and ROSAT PSPC data.  Soft excess in Abell~1795 was also found
with the ROSAT PSPC \citep{bonamente2003} at the $10-20$~\% level in the
$0.2-0.4$~keV band, and by \citet{nevalainen2003} and \citet{kaastra2003}, both based on XMM-Newton data (see Sect.~4.2.1 and
4.2.2, respectively).

\subsubsection{Abell~2199 ($z=0.0301$) }
\label{a2199}

The presence of a soft excess was disputed in Abell~2199. From EUVE
data, \citet{lieu1999b} detected a soft excess, and confirmed its
existence up to a radius of 20$^{\prime}$ using a background measured from
offset pointing \citet{lieu1999c}. Confirmation of the existence of
a soft and a hard excess in this cluster was given by \citet{kaastra1999} based on BeppoSAX, EUVE and ROSAT data. A three phase ICM was
proposed to account for the observations by \citet{lieu2000}.  

On the other hand, from other EUVE observations taking into account the
variations of telescope sensitivity over the field of view, \citet{bowyer1999} found no evidence for EUV excess. They confirmed this
result with BeppoSAX data \citep{berghofer2002}, but \citet{kaastra2002} claimed that their analysis was wrong. According to the
latter authors, the problem was that Bergh\"ofer \& Bowyer based their
conclusion on a plot of the ratio of the observed radial intensity
profiles in the $0.5-2.2$~keV band as compared to the $0.1-0.3$~keV
band. Because the point spread function of the BeppoSAX LECS
instrument is a strong function of energy (10$^\prime$ FWHM
at $E=0.28$~keV, scaling as $E^{-0.5}$), and the low-energy counts in
the $0.1-0.3$~keV band contain a significant contribution from higher
energy photons, even for a fully isothermal cluster the ratio of the
radial intensities in both bands shows strong ($\sim$50~\%)
excursions. Moreover, the predicted ratio depends strongly on issues
like abundance gradients, modelling of the cooling flow, etc. that were
ignored by \citet{berghofer2002}.

The existence of a soft excess was independently confirmed by \citet{durret2002} from EUVE and ROSAT PSPC data. The ROSAT PSPC analysis
of \citet{bonamente2003} also revealed marginal evidence for soft
excess in Abell~2199, and they derived that the temperature of the
soft excess component must be below 0.2 keV. \citet{kaastra2003}
confirmed the presence of a soft excess in this cluster from
XMM-Newton data.

\subsubsection{Abell~4059 ($z=0.0460$)}
\label{a4059}

No soft excess was detected in Abell~4059 by \citet{berghofer2000b}; strangely, these authors even found a deficit of EUV
emission in the central $2^\prime$. However, a soft excess was detected
in this cluster by \citet{durret2002} as described in the next
subsection.

\subsubsection{Complementary results on the five above clusters}
\label{complem}

\begin{figure}    
\begin{center}
\includegraphics[width=\textwidth]{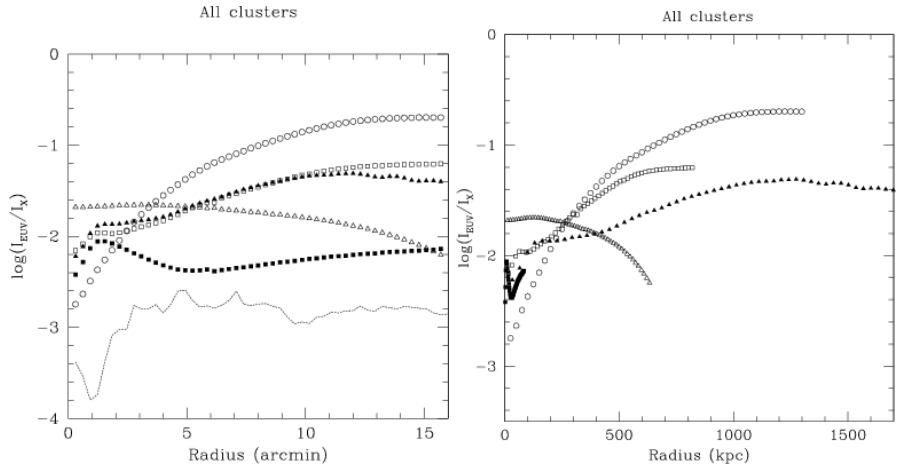}
\caption{EUV to X-ray ratio as a function of cluster radius (expressed
in arcmin in the left plot and in kpc in the right plot) for the five
clusters observed by \protect\citet{durret2002} with the EUVE and ROSAT PSPC
satellites. The symbols are the following: Abell~1795: filled
triangles, Abell~2199: empty squares, Abell~4059: empty circles, Coma:
empty triangles, Virgo: filled squares. The dashed line in the left
figure shows the ratio of the EUVE to ROSAT PSPC PSFs. Error bars are
omitted for clarity; they are typically smaller than $\pm 0.01$ in
logarithmic scale, i.e. too small to be clearly visible on the figure.}
\label{fig_durret02}
\end{center}  
\end{figure}

A different approach was proposed by \citet{durret2002} based on the
wavelet analysis and reconstruction of EUVE and ROSAT PSPC observations 
for the five
previously discussed clusters: Virgo, Coma, Abell~1795, Abell~2199 and
Abell~4059. A soft excess was found in all five clusters, even when
taking into account temperature and abundance gradients of the ICM.
The radial profiles of the EUV to X-ray ratios are shown in
Fig.~\ref{fig_durret02}.  The EUV and X-ray profiles were shown to
differ statistically, suggesting that the EUV and X-ray emissions were
probably not due to the same physical mechanism.

\subsubsection{Fornax ($z=0.0046$)}
\label{fornax}

No soft excess was detected in Fornax by \citet{bowyer2001} at any
scale, but \citet{bonamente2002} did detect a soft X-ray excess
in this cluster, which is part of the sample described in Sect.~3.2.9.

\subsubsection{Shapley supercluster ($z=0.046$)}

The ROSAT PSPC data of four clusters Abell~3571, Abell~3558,
Abell~3560 and Abell~3562 in the Shapley supercluster were analysed by
\citet{bonamente2001b}.  They modelled the hot gas spectra using
published ASCA results (Abell~3571 and Abell~3558) or fitting the PSPC
data (Abell~3560 and Abell~3562).  These models fit the PSPC data
poorly due to excess residuals in the R2 band ($0.15-0.3$ keV) in all
clusters. Based on the residuals, the authors determined the
fractional soft excess emission above the hot gas emission model,
amounting to 50\% at most.  For Abell~3571 and Abell~3558, they found
that the fractional soft excess increased with the cluster radius, an
effect already found in the EUVE data of Virgo, Abell~1795, and
Abell~2199 \citep{bonamente2001b,lieu1999b}.

BeppoSAX LECS data on Abell~3571 and Abell~3562 supported the PSPC results.
The analysis showed that a simple one-temperature thin-plasma model is
not adequate to describe the emission.

The data indicated a thermal model with ${\mathrm k}T=0.1$~keV for Abell~3558, but the
quality of the data was not appropriate for distinguishing the nature
of the soft component or modelling its properties in detail.

\subsubsection{S\'ersic 159-03  ($z=0.0580$)}
\label{sersic159}

S\'ersic~159-03 (also known as AS\,1101) was found to show strong soft excess and even to be
the brightest soft excess cluster by \citet{bonamente2001d}.  These
authors combined three PSPC pointings of S\'ersic~159-03 obtaining
19~ks of exposure time.  The low temperature ($\sim$2~keV) of the hot
gas allowed its spectroscopic analysis using the PSPC data alone.
Extrapolating the best-fit spectral model from the $0.5-2.0$~keV band
to lower energies revealed a very strong soft excess at 0.2~keV, 100~\%
over the hot gas emission. This excess was then confirmed with
XMM-Newton (see Sect.~4.1.3).

\citet{bonamente2001d} tested the hypothesis that the soft excess
could be an artefact caused by a sub-Galactic absorbing column density
$N_{\rm H}$ towards the cluster.
However, this yielded an unrealistically low $N_{\rm H}$ ($5
\times 10^{19}$~cm$^{-2}$), even lower than that of the global
minimum of the Lockman Hole, implying that $N_{\rm H}$ variation is not
the cause of the soft excess in this cluster.

Despite the high quality of the signal, the nature of the soft
excess component could not be determined, since
non-thermal and thermal models yield statistically equally acceptable
fits.

\subsubsection{The Hercules supercluster ($z\sim 0.035$)}
\label{hercules}

A soft excess was also detected from low energy ROSAT data in the
clusters Abell~2052 ($z=0.0348$) and Abell~2063 ($z=0.0348$) of the
Hercules supercluster \citep{bonamente2005b}; however these authors
underlined the difficulty of background subtraction.

\subsubsection{The Bonamente et al. (2002) sample}

\citet{bonamente2002} published a ROSAT PSPC analysis of the largest
soft excess sample of galaxy clusters so far, 38 in number. They
modelled the ROSAT PSPC data of the hot gas using published results
from ASCA and BeppoSAX which are more suitable due to their energy
coverage up to 10 keV. These models were extrapolated to the PSPC
channels and compared with the data in the $0.2-0.4$~keV band. The
analysis yielded significant (3$\sigma$) soft excess in $\sim$30~\% of
the clusters in the sample. The actual fraction of clusters with soft
excess may be higher, since the clusters with the deepest ROSAT
observations all had soft excess.  Using a thermal model with a
temperature of 0.08 keV and metal abundance of 0.3 Solar, or a
power-law $I = I_0  E^{-\alpha _{\mathrm{ph}}}$ (where $I$ is the intensity
and $E$ the photon energy) of index $\alpha_{\mathrm{ph}} = 1.75$, the soft
excess fluxes correspond to $0.2-0.4$~keV luminosities of $10^{40} -
10^{43}$ erg\,s$^{-1}$. The excess was more prominent in the regions
outside the central $150-200$~kpc.

The indicated increase of the relative strength of the soft excess
with radius is qualitatively consistent with the non-thermal Inverse
Compton radiation of CMB photons from the relativistic cluster
electrons.
The hot gas emission is proportional to $n_{{\rm e,thermal}}^2$ while the
IC is proportional to $n_{{\rm e,relativistic}}$ and the usual assumption
is that the relativistic electrons follow the gas distribution.
This model however faces energetic problems.

The authors also examined the possibility that the soft excess
originates from a large number of unresolved X-ray-emitting cluster
galaxies.  The most relevant sources of X-rays are Low Mass X-ray
Binaries (LMXB) and especially their ${\mathrm k}T\sim 0.25$ keV blackbody
component coming from the accretion disk.  The typical soft X-ray
luminosity of a galaxy due to LMXB is of order $10^{38}$~erg~s$^{-1}$
and thus $10^{3} - 10^{4}$ galaxies are needed to produce the typical
observed soft X-ray luminosity, while typical numbers of galaxies in
rich clusters are of order several hundreds. Thus, unresolved sources
in cluster galaxies cannot explain the soft excess. A similar
conclusion can be reached by looking at the galaxy X-ray luminosity
function, such as that obtained for the Coma cluster by \citet{finoguenov2004}; in view of the relatively faint X-ray luminosities of the
individual galaxies, the number of X-ray emitting galaxies required to
account for the soft excess would be unrealistically high.

\section{Soft X-ray emission based on XMM-Newton EPIC data}
\label{xmm}

\subsection{Continuum detections}

\subsubsection{The Nevalainen et al. (2003) sample}

\citet{nevalainen2003} found evidence for soft X-ray excess in
clusters of galaxies using XMM-Newton EPIC data.  They derived a
$20-40$~\% soft excess in the central 500~kpc regions of the clusters
Coma, Abell~1795 and Abell~3112 ($z=0.0703$) in channels below 2 keV,
consistently in both PN and MOS instruments, and in ROSAT PSPC data.

A thermal model fits the data better than a non-thermal one, but at
the level of calibration accuracy at the time, the non-thermal nature
of the soft excess could not be ruled out.

Thermal modelling yielded temperatures in the range of $0.6-1.3$~keV and
metal abundances consistent with zero.
Assuming that this warm gas occupies the same volume as the hot gas,
the electron densities are of the order of $10^{-4} - 10^{-3}$~cm$^{-3}$. These values lead to a cooling time scale larger than the
Hubble time, i.e. the structures are self-consistent.

\subsubsection{The Kaastra et al. (2003) sample}

\citet{kaastra2003} examined a sample of 14 clusters of galaxies
observed with XMM-Newton, and found significant evidence for soft
X-ray excess continuum emission in 5 of them: Coma (centre studied
only, since it is closer than the other 4), Abell~1795,
S\'ersic~159-03, Abell~2052 and MKW~3s. Different modelling of the hot
gas temperature and metal abundance yielded lower temperatures for the
soft component ($\sim$0.2~keV), compared to \citet{nevalainen2003}. The surface brightness of the warm gas is rather constant
with radius, while that of the hot gas decreases with radius, falling
below the warm gas surface brightness between 0.5 and 1~Mpc from the cluster
centre.  
Note that in Coma, the central pointing does not cover this radius, so
the hot gas brightness remains above the cool gas brightness, consistent with the
above number.
Such a behaviour is consistent with the Warm Hot Intergalactic
Medium (hereafter WHIM) filament scenario, whereby the projected
external filamentary structure is more extended than the cluster.
Later, \citet{kaastra2004} extended this sample to 21 clusters, and
found 7 objects with soft excesses, the new cases being Abell~3112
(also in Nevalainen's paper) and Abell~2199.

Based on the quality of the spectral fits, in most cases the
non-thermal model was also acceptable.  Fitting the soft excess with a
power-law model yielded rather constant photon indices ($\sim$ 2) with
radius in a given cluster. At the cluster centres the luminosity of
the non-thermal component is $\sim 10\%$ of that of the hot gas, while
the percentage increases towards 100\% at the largest radii.

\citet{bregman2006} challenged the results of \citet{kaastra2003}, arguing that their soft excess detection was due to
incorrect background subtraction. However, in a rebuttal paper,
\citet{nevalainen2007} showed that, especially in the central
regions, the cluster emission is so bright compared to the background,
that the details of the background modelling are insignificant.  Thus,
Bregman \& Lloyd-Davies's claim appears to be unjustified.

\citet{nevalainen2007} also found that the changes in the EPIC
calibration between the years 2002 and 2005 resulted in a decrease of
the soft excess signal.  Using the \citet{kaastra2003} modelling, the
PN soft excess even disappeared in some clusters. However, the MOS
instrument still detects a soft excess in all the clusters of
Kaastra's sample. A less conservative hot gas modelling (Nevalainen et
al., in preparation) with the current calibration information obtains
consistent soft excesses in both PN and MOS data.

\subsubsection{S\'ersic 159-03}

One of the first clusters to be observed by XMM-Newton was S\'ersic
159-03. This observation with a net exposure time of 30~ks was taken in
2000 \citep{kaastra2001}. The same observation was used for a search
for soft excess emission (\citealt{kaastra2003}, see Sect.~4.1.2
above). A 60~ks observation taken two years later was analysed
by \citet{bonamente2005a} and \citet{deplaa2006}.

\citet{bonamente2005a} confirmed the existence of the soft X-ray
excess emission in S\'ersic~159-03 out to a distance of 1 Mpc from the
cluster centre. The soft excess in the $0.3-1.0$ keV band increases
from 10~\% at the centre to 80~\% at the largest radii. The properties of
the soft excess differ from those derived for S\'ersic~159-03 using
PSPC data \citep{bonamente2001d}, likely due to different modelling
of the data: XMM-Newton allows to determine the hot gas component
unambiguously in the $2.0-7.0$~keV band where the soft component has a
negligible contribution.

The same dataset of S\'ersic 159-03 was also analysed by \citet{deplaa2006}. However, they focused their attention on the chemical
evolution of the cluster and they determined the background for the
spectral analysis in the $9^\prime-12^\prime$  region around the cluster,
subtracting the spatially extended soft excess emission together with
the soft foreground emission. For the remaining soft excess in the
cluster core they propose a non-thermal emission mechanism arising
from IC scattering between CMB photons and relativistic electrons
accelerated in bow shocks associated with ram pressure stripping of
infalling galaxies.

\citet{werner2007} studied Suzaku data together with the two XMM-Newton data
sets of S\'ersic 159-03 obtained two years apart mentioned
before. They found consistent soft excess fluxes with all instruments
in all observations. From the XMM-Newton data they derived radial profiles
and 2D maps that show that the soft excess emission has a strong peak
at the position of the central cD galaxy and does not show any
significant azimuthal variations. They concluded that the spatial
distribution of the soft excess is neither consistent with the models
of intercluster warm-hot filaments, nor with models of clumpy warm
intracluster gas associated with infalling groups as proposed by
\citet{bonamente2005a}. Using the data obtained with the XMM-Newton
RGS, Werner et al. could not confirm the presence of warm gas in the
cluster centre with the expected properties assuming the soft excess
was of thermal origin. They therefore concluded that the soft excess
in S\'ersic 159-03 is most probably of non-thermal origin.

\subsection{Line emission detection}

A crucial piece of evidence for the thermal nature of the soft X-ray
excess would be the detection of emission lines.  At the indicated
temperatures of $0.1-0.5$~keV the most prominent emission lines are from \ion{O}{vii} and \ion{O}{viii}.  The resonance, intercombination and forbidden lines
of \ion{O}{vii} have energies of 574, 569 and 561 eV (see also \citealt{kaastra2008} - Chapter 9, this volume).  For a low density
plasma at 0.2~keV temperature, the centroid of the triplet has an
energy of 568.7 eV (see \citealt{kaastra2003}).

Geocoronal and heliospheric solar wind charge exchange \citep{wargelin2004,fujimoto2007} also produces soft X-ray emission
lines. These lines can vary by a factor of 3 on time scales of hours,
which makes them difficult to model and subtract. The maximum observed
brightness of these lines in the $0.5-0.9$ keV range can reach the level
of the cosmic X-ray background. Therefore, when the background level
is important for the analysis, care should be taken when estimating
the contaminating effects of this emission.

\citet{finoguenov2003} observed a spatial variation of the soft
excess emission in the Coma cluster. In the Coma 11 field they found
an excess emission which is particularly strong. Recent observations
with the Suzaku satellite (Takei et al., private communication) do not
confirm the level of \ion{O}{vii} and \ion{O}{viii} line emission reported by
\citet{finoguenov2003} for the Coma 11 field, but they are
consistent with the lowest reported values for the other fields in
Coma observed with XMM-Newton.
They suggest that a large fraction of the reported excess soft X-ray
emission and the line emission observed in the Coma 11 field with
XMM-Newton was due to Solar wind charge exchange emission.

\subsubsection{The Kaastra et al. (2003) sample}

\citet{kaastra2003} found evidence for \ion{O}{vii} line emission in the
clusters S\'ersic~159-03, MKW~3s ($z=0.0450$) and Abell~2052
in form of line-like residuals on top of the hot gas
model at wavelengths consistent with the cluster redshifts.  Note that
the line emission is significant only in the outer regions ($4^\prime-12^\prime$) of the clusters.

The uncertainties also allow a Galactic origin ($z=0$) for the
emission.  Also, \citet{nevalainen2007} pointed out that a simple
cluster-to-background flux level comparison does not exclude that the
reported \ion{O}{vii} line emission in the $0.5-0.65$ keV band contains
contributions from the geocoronal and heliospheric Solar wind charge
exchange (see above). This possibility remains to be studied in
detail.

Therefore this study does not give conclusive evidence for \ion{O}{vii} line
emission in these three clusters.

\subsubsection{Coma}

\begin{figure}    
\begin{center}
\includegraphics[width=0.6\textwidth]{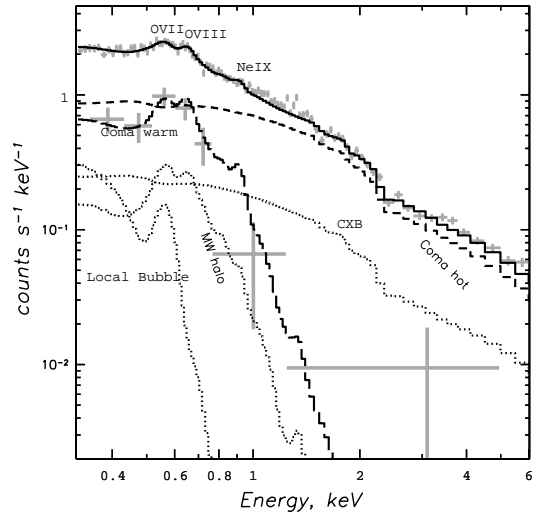}
\caption{Different emission components of the Coma spectrum
obtained with the XMM-Newton PN instrument (from \protect\citealt{finoguenov2003}).}
\label{fin.fig}
\end{center}  
\end{figure}

\citet{finoguenov2003} analysed XMM-Newton data in the outskirts of
the Coma cluster.  The spectrum at $\sim$1~Mpc distance from the
centre exhibits a very strong, $\sim$100~\% fractional soft excess
continuum in channels below 0.8 keV, and the authors detect
two emission lines in the $0.5-0.6$~keV band.  Both the excess
continuum and the line features are well fit with a thermal model of
${\mathrm k}T = 0.22$~keV (see Fig.~\ref{fin.fig}).  The temperature, as well as
the derived baryon overdensity for the warm gas ($\sim$200) are
consistent with the WHIM filament properties in numerical simulations
(e.g. \citealt{dave2001,yoshikawa2006}).
Let us note however that the soft excess in Finoguenov's model consists
mostly of line emission, but due to the low spectral resolution of
EPIC at low energy, it looks more like a continuum.

The redshift of the \ion{O}{vii} line ($0.0-0.026$) is consistent with that of
Coma (0.023) and also with being Galactic (0.0), implying that the
emitter is located between us and Coma.  Optical data show that there
is a significant galaxy concentration in the direction of the warm gas
in front of Coma.  This suggests that the warm gas and the galaxy
structure are connected, consistent with the WHIM filament scenario.

However, since the soft excess flux in Coma is comparable to the
background level, the charge transfer flux in an active state (i.e. at
the maximum observed level) can reach the soft excess level detected
in Coma \citep{finoguenov2003}. \citet{bowyer2004} use this
coincidence as a proof that the soft excess detection in Coma is due
to charge transfer.  However, they did not model the soft excess
spectrum with a charge transfer emission model to show that the model
is consistent with the observed spectral features in Coma. Thus, their
claim of Coma soft excess being due to charge transfer mechanism is not
proven.

An indication for the presence of WHIM in Coma was also recently
reported by \citet{takei2007b} at 2.3$\sigma$ level in absorption
and in emission, based on RGS spectra of the Active Galactic Nucleus
X--Comae, which is located behind the Coma cluster.

\section{Soft X-ray emission based on Suzaku observations}
\label{Suzaku}

\subsection{Search for the soft excess in and around clusters}

Suzaku has observed about 30 clusters of galaxies by April
2007. Several of these observations were done in order to search for
soft excess emission. We describe here the results for Abell~2218 and
Abell~2052 in some detail since these two clusters have been studied
rather extensively, and then briefly review results from other
sources.

\subsection{Abell 2218 ($z=0.1756$)}

Abell~2218 is a bright cluster at $z=0.1756$ with an ICM temperature
of about 7 keV. The Suzaku observations were carried out on two
occasions with a total exposure of 80 ks and the results were
published by \citet{takei2007a}. Several conditions are favourable
to the detection of a soft excess in this cluster with Suzaku. First,
the cluster is known to be undergoing a merger, likely along the line
of sight, as indicated by two galaxy concentrations at different
redshifts. This picture is supported by X-ray observations from
Chandra and by the presence of remarkable arcs due to gravitational
lensing. Such a merger suggests that a large-scale filament may be
located along the line of sight and that the column density of oxygen
contained in the warm filament gas could be high. The second reason is
that the XIS energy resolution allows to separate a $\sim 100$~eV
cosmologically redshifted oxygen line from a zero redshift oxygen
line.  Background data in two regions at about $5^\circ$ offset from
the cluster were also taken to obtain information on the oxygen line
intensity of the Galactic component.

The time variation of the soft X-ray flux during the observations was
very small, so the effect from the solar wind charge exchange emission
was considered to be small. A soft excess was searched for in the
outside region of the cluster, i.e.\ for the region with radius
greater than $5^\prime$ (880~kpc at the source) from the Abell~2218
centre. The observed spectra (see Fig.~4) show structures around 0.5
and 0.6 keV, which are the energies corresponding to redshifted \ion{O}{vii}
and \ion{O}{viii} lines. However, close examination showed that the hump at
0.5 keV can be caused by the oxygen edge in the XIS filters, and the
peak around 0.6 keV can be due to Galactic \ion{O}{vii} line emission. This
unfortunate situation hampered the original XIS capability of
detecting the soft excess associated with the cluster. \citet{takei2007a} set upper limits for the \ion{O}{vii} and \ion{O}{viii} line intensities
as shown in their Fig.~1, including all the uncertainties in the
instrumental response (e.g. by the enhanced contamination in the XIS
filters) and in the Galactic line intensities.

Even though the observing conditions were not optimal, the upper
limits obtained are nearly an order of magnitude lower than the soft
excess level reported in other clusters by \citet{kaastra2003}. With this upper limit, the density of the warm gas can be
constrained. The limit for the gas density is:
\begin{equation}
 \delta \equiv n_{\rm H}/\bar{n}_{\rm H} < 270~
  \left(\frac{Z}{0.1~Z_\mathrm{\odot}}\right)^{-1/2}~
  \left(\frac{L}{2~\mathrm{Mpc}}\right)^{-1/2},
\end{equation}
where $\bar{n}_{\rm H} = X\Omega_\mathrm{b}\rho_\mathrm{crit} (1+z)^3
/m_\mathrm{p} = 1.77\times10^{-7}~(1+z)^3~\mathrm{cm^{-3}}$ is the
mean hydrogen density in the universe, where $X=0.71$ is the
hydrogen to total baryon mass ratio, $\Omega_\mathrm{b}=0.0457$ is the
baryon density of the universe,
$\rho_\mathrm{crit}=9.21\times10^{-30}~\mathrm{g~cm^{-3}}$ is the
critical density of the universe, and $m_\mathrm{p}$ is the proton
mass. Even though this level of gas density is much higher than the
expected density ($\delta \sim 10$) in the filaments, the result
demonstrates that Suzaku can give a reasonable constraint on the
soft excess emission.

\subsection{Abell 2052}

Suzaku carried out observations of Abell~2052 with 4 offset pointings
to cover the cluster outskirts to about $20^\prime$ (830~kpc at the source).
Since the observations were all carried out only 40 days after the
launch, the contamination in the XIS filter was not so much a
problem. At the carbon K-edge energy, the reduction of the
transmission efficiency was less than 8~\% (see \citealt{koyama2007}).

The data were analysed by \citet{tamura2007}. By assuming a single
temperature for the ICM and two temperatures (0.2 and 0.6 keV) for the
Galactic component, they found a significant soft excess, with a
spectrum well described by a featureless continuum modelled (for
convenience reasons only) by pure thermal Bremsstrahlung with a
temperature around 0.7 keV. The intensity of this soft component is
consistent with a constant value over the entire observed field and
stronger than the Galactic emission below 0.5 keV.  If this emission
is associated with Abell~2052, then its luminosity can be comparable
to the ICM emission which shows $L_X \sim 1.4 \times 10^{44}$ erg\,
s$^{-1}$ in the $2-10$ keV band. The cluster is located fairly close to
the extension of the North Polar Spur (NPS), a large spur-like region
with strong soft X-ray emission (see e.g. \citealt{willingale2003}).
However, the Suzaku spectrum of the NPS region showed the temperature
to be about 0.3 keV with a fairly strong Mg-K emission line. This is
inconsistent with the Abell~2052 soft excess spectrum.

A strong soft component without emission lines may be caused by
extended non-thermal emission. Since the cluster shows neither strong
radio emission nor merger features, such a non-thermal emission could
be due to a rather old population of non-thermal electrons
(see e.g. \citealt{rephaeli2008} - Chapter 5, this volume).

\subsection{Other clusters}

The Sculptor supercluster was observed with Suzaku in 4 pointings with
the fields partially overlapping and connecting 3 main clusters
(Abell~2811 at $z=0.1086$, Abell~2804 at $z=0.11245$) and Abell~2801
at $z=0.11259$). The data were analysed by \citet{kelley2007}. The
combined spectrum after removing bright clusters and point sources
shows an excess in the energy range $0.6-1$~keV, with a temperature of
about 0.8 keV. Since the observed volume is extremely large ($\sim
500$ Mpc$^3$), the uniform electron density implied is as low as
$8\times 10^{-6}$~cm$^{-3}$.

The Suzaku XIS data on Abell~1060 ($z=0.01140$, \citealt{sato2007a}),
AWM7 ($z=0.01724$, \citealt{sato2007b}), and Fornax \citep{matsushita2007} have been analysed in some detail. Even though the spectrum and
intensity of the foreground Galactic emission have fairly large
ambiguities, the observed energy spectra for these clusters are
generally well fit by thermal models at energies down to about 0.3
keV. For the central region of Fornax ($r<2^\prime$), a two temperature
model with ${\mathrm k}T = 1.5$ and 0.8 keV was preferred. However, excess
emission which causes significant deviation from these standard
spectral fits has not been detected from these clusters.

\begin{figure}
\hbox{\includegraphics[width=0.49\hsize]{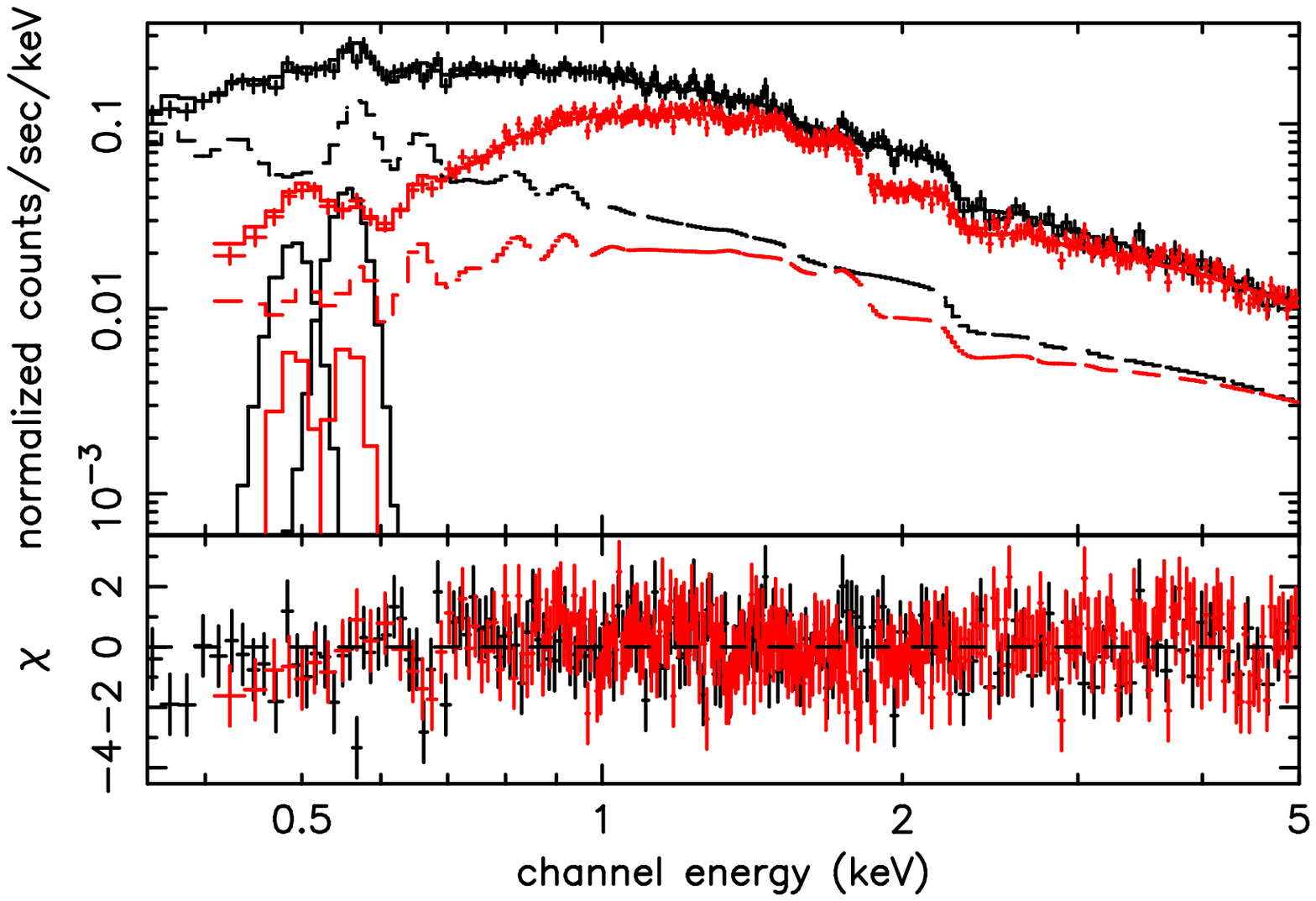}
\includegraphics[width=0.49\hsize]{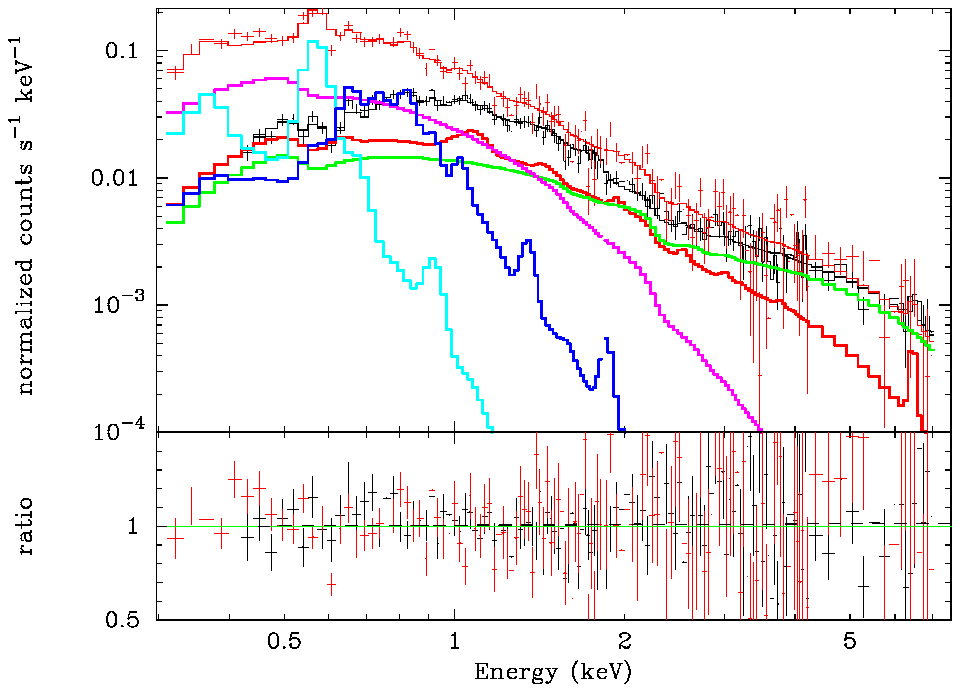}
}
 
\caption{\emph{Left:} Observed XIS spectrum for the outer region of
Abell~2218, with upper limits for the redshifted \ion{O}{vii} and \ion{O}{viii} lines,
for the BI (black) and FI (red) sensors. The background spectra are
shown with dashed lines. This figure is taken from \protect\citet{takei2007a}.
\emph{Right:} Observed XIS spectrum for $r=12^\prime-15^\prime$. Spectral fit including
Galactic emission (0.1 and 0.3 keV in light and dark blue), cosmic
X-ray background (green), and thermal ICM emission with ${\mathrm k}T=2.8$ keV
(red) requires an additional thermal Bremsstrahlung component with
${\mathrm k}T=0.7$~keV (magenta) whose intensity is constant over $r=0-20^\prime$. The
summed up model (orange) is fitted to the data.  Observations with
Suzaku confirm the presence of soft excess emission in S\'ersic 159-03
and its derived flux is consistent with the values determined with
XMM-Newton (Sect.~4.1.3, \protect\citealt{werner2007}). However, Suzaku does not
confirm the presence of the redshifted \ion{O}{vii} lines in the cluster. The
excess emission can be fit statistically equally well with a thermal
model with low oxygen abundance ($<0.15$ Solar) and with a non-thermal
model.  The figure is taken from \protect\citet{tamura2007}.}
\end{figure}

\citet{fujita2007} observed the region between Abell~399 and
Abell~401 with Suzaku.
They found no evidence for oxygen emission from the WHIM in
this region and obtained a strict upper limit of $4.1\times
10^{-5}$~cm$^{-3}$ on its density.

\section{Soft X-ray emission based on Chandra observations}

Reports on soft excess emission in clusters as observed by Chandra are
scarce.  \citet{henriksen2004} detected soft excess emission in the
merging cluster Abell~754 ($z=0.0535$).  Although a non-thermal origin
could not be fully excluded, from their combined Chandra, BeppoSAX,
ASCA and ROSAT PSPC analysis they conclude that a weak soft thermal
component ($\sim 1$~keV) is present in this hot ($\sim$ 10~keV)
cluster. The emission extends out to 8$^{\prime}$ from the core and is
peaked in the cluster centre. Henriksen et al. attribute the emission
to embedded groups of galaxies.

In another cluster, Abell~2163 ($z=0.2030$), \citet{henriksenhudson2004} also report a soft and hard X-ray excess in their combined
Chandra and ROSAT PSPC analysis. The excess can be modelled by a
non-thermal component with photon index $2.7 - 5.9$. Little detail
about the observations is given in this paper, but the authors
attribute the non-thermal emission to a weak merger shock that is
present in this cluster.

\citet{bonamente2007} analysed Chandra data of Abell~3112 in the
central $1.0-2.5$ arcmin region and found significant soft excess,
reaching 20~\% of the hot gas emission level at 0.3~keV.  Its spectrum
is acceptably fit with a low temperature ($\sim$0.5~keV) thermal
model or by a non-thermal model ($\alpha_{\mathrm{ph}} \sim 1.8$).  The best-fit
values differ from those obtained with XMM-Newton data, indicating
that remaining calibration uncertainties affect fine details.
Nevertheless, the magnitude of the soft excess is above the level of
calibration inaccuracies, and similar to that in XMM-Newton, thus
proving the celestial origin of the soft excess in Abell~3112.

\section{Discussion}
\label{discussion}
Soft excess emission, defined here as excess emission over the usual
thermal emission from hot intracluster gas at energies below 1~keV,
has been detected significantly in $\sim$30~\% of clusters of galaxies.
This value is based on the two largest cluster soft excess samples
\citep{bonamente2003,kaastra2003}.  However, as described
above, such detections are difficult because of instrumental issues
and for many objects a controversy remains.

Thermal and non-thermal models have been proposed to account for this
soft excess emission. We will briefly summarise them below using Coma
and S\'ersic 159-03 as examples.

\subsection{Thermal models}
\label{thermal}

\subsubsection{WHIM in Coma}

The massive soft excess halo around Coma \citep{bonamente2003} is
not well accounted for by a non-thermal model.  Furthermore,
\citet{finoguenov2003} detected \ion{O}{vii} and \ion{O}{viii} line emission in the
outskirts of Coma.  Thus Coma is a good case for a cluster where the
soft excess is dominantly of thermal nature.  \citet{bonamente2003}
proposed that the soft component may reside in external filamentary
structures of warm-hot intergalactic medium (WHIM). These filaments
are predicted by hydrodynamic simulations of formation and evolution
of large-scale structures, extending for several megaparsecs,
containing a large fraction of the current epoch's baryons (e.g. \citealt{cen1999,dave2001,yoshikawa2006}).  In
WHIM simulations, the temperatures are in the range $10^{5}-10^{7}$~K,
consistent with the soft excess properties found in Coma.  Assuming a
filamentary geometry, \citet{bonamente2003} derived densities of
$n\sim 10^{-5}-10^{-4}$~cm$^{-3}$ for the warm gas in Coma.  These
values are consistent with those in the WHIM simulations ($10^{-7}-
10^{-4}$~cm$^{-3}$).  Assuming $n=10^{-4}$ cm$^{-3}$, the implied
mass of the filament exceeds that of the hot gas by a factor of 3.

\subsubsection{Soft emission from merging sub-halos}

In S\'ersic~159-03, detailed WHIM filament calculations showed that
unrealistically long structures ($\sim 100 - 1000$ Mpc), projected on
the line of sight of the cluster, are required to explain the soft
excess detected in XMM-Newton data \citep{bonamente2005a}.  The
assumption that the warm gas occupies the same volume as the hot gas
has the problem that there is a pressure difference between the two
components, and that the cooling time is short. As proposed by these
authors, this problem can be avoided if one assumes that the warm gas
is not distributed evenly but rather in high density $10^{-2} -
10^{-3}$ cm$^{-3}$ clumps with a volume filling factor $f\ll 1$.  Such a
distribution is predicted by the simulations of \citet{cheng2005},
in which the soft excess emission comes from high-density and low
entropy gas associated with merging sub-groups, which preserve their
identity before being destroyed and thermalised in the hot ICM. The
simulated 0.2--1.0 keV band soft excess in the radial range of 
$0-0.5 r_{\mathrm{vir}}$ is consistent with that published in \citet{bonamente2005a} for S\'ersic~159-03.  This scenario yields a warm gas mass
of 25~\% of the hot gas mass for S\'ersic~159-03.

However, \citet{werner2007} show that the soft excess emission peaks
at the position of the central cD galaxy and does not show any
significant azimuthal variations. Moreover the soft excess in
S\'ersic~159-03 is observed out to radii of at least 1~Mpc. If this
soft excess is associated with the gas of an infalling group, then
this group is moving exactly along the line of sight. However, such an
infalling group cannot explain the presence of the soft excess
emission at large radii. Therefore the soft excess observed in
S\'ersic~159-03 is most probably not of thermal origin.

\subsection{Non-thermal models}
\label{non-thermal}

Inverse Compton models of energetic electrons on CMB
photons and/or on galaxy starlight have been developed by various
authors to account for the soft excess observed in several of the
objects presented here.

Non-thermal inverse Compton emission has a power-law spectrum with a
relative flux which in the $0.3-10.0$~keV band may account for more
than 30~\% of the cluster emission.  The best-fit power-law photon
indices of soft excess clusters are typically between $\alpha_{\mathrm{ph}}
\sim 2.0-2.5$.  In the IC model, this corresponds to a differential
relativistic electron number distribution ${\mathrm d}N/{\mathrm d}E = N_0 \,
E^{(-\mu)}$ with $\mu = 3-4$.  These are steeper than the distribution
of the Galactic cosmic-ray electrons ($\mu\sim 2.7$). The steeper
power-law distribution might indicate that the relativistic electrons
suffered radiative losses \citep{sarazin1999}. Relativistic electrons in
this energy range have relatively long lifetimes of $t_{\mathrm{IC}}
=2.3\times10^9 (\gamma/10^3)^{-1}(1+z)^{-4}$~yr and
$t_{\mathrm{syn}}=2.4\times10^{10}(\gamma/10^3)^{-1}(B/(1\mu
G))^{-2}$~yr for inverse-Compton and synchrotron processes
respectively.

As mentioned earlier, \citet{werner2007} conclude that a non-thermal
model best explains the observed properties of the soft excess in
S{\'e}rsic~159-03. The total energy in relativistic electrons needed
to explain the excess emission within the radius of 600~kpc does not
exceed 1$\times 10^{61}$~erg, while the total thermal energy within
the same radius is 3$\times 10^{63}$~erg. This means that even if the
energy in relativistic ions is as much as $\sim$30 times larger than
that in relativistic electrons, the total energy in cosmic ray
particles will only account for 10~\% of the thermal energy of the ICM.

Models that may account for the observed soft excess can also be found
in the following list: \citet{ensslin1999,atoyan2000,sarazin2000,takizawa2000,fujita2001,petrosian2001,depaolis2003,bowyer2004a,petrosian2008} - Chapter 10, this volume.

\subsection{Some problems and open questions}
\label{problems}

In the Virgo cluster a strong soft excess was detected in the extreme
ultraviolet with EUVE \citep{lieu1996a,berghofer2000a,bonamente2001b,durret2002}
and in the $0.2-0.4$~keV
band with ROSAT \citep{bonamente2002}.  However, the observations
with XMM-Newton did not confirm the existence of this soft excess: a
thermal model for the hot cluster emission with Galactic absorption
describes the soft band X-ray spectra (above 0.3 keV) of Virgo
obtained with XMM-Newton sufficiently well \citep{kaastra2003,matsushita2002}.

\citet{arabadjis1999} claimed that some of the X-ray absorption
cross sections were wrong, and that soft excesses would disappear when
using the proper values. However, it is surprising that this claim has
neither been confirmed nor refuted ever since.  During some time,
there have been wrong He cross sections by \citet{balucinska1992} in the XSPEC software (see
http://heasarc.gsfc.nasa.gov/docs/xanadu/xspec/). Arabadjis \& Bregman
refer to those wrong cross sections. They have been improved by \citet{yan1998} and are now included properly for example in Wilms' cross
sections that are in XSPEC. They are also properly included in the
SPEX software. Note that Wilms' more recent cross sections agree quite
well with the older \citet{morrison1983} work  (see the
discussion in \citealt{wilms2000}).  Lesson to be learned: it is
important to check for each paper which cross sections / absorption
model have been used!

Several searches for far-ultraviolet emission lines expected from a
$10^6$~K gas were performed with the FUSE satellite on the cores of
several clusters. \citet{oegerle2001} have reported the detection of
\ion{O}{vi} $\lambda$1032~\AA\ in Abell~2597 ($z=0.0824$), implying a mass inflow 
rate of about 40~M$_\odot$/yr. However, FUSE has not detected warm gas
in five other clusters: Abell~1795 \citep{oegerle2001}, Coma and
Virgo \citep{dixon2001}, and Abell~2029 ($z=0.0775$) and
Abell~3112 \citep{lecavelierdesetangs2004}; the upper limits on
the inflow rates for these five clusters were of the order
25~M$_\odot$/yr.

A recent review by \citet{bregman2007} summarises current knowledge on the
search for missing baryons at low redshift. Note however that in his
Fig.~9 (already in \citealt{bregman2006}) he claims that the
soft excess found in some clusters by \citet{kaastra2003} is an
artefact due to wrong background subtraction. The argument is that
there is a correlation between the Rosat R12 (low energy) count rate
and the presence or absence of a soft excess. However, a closer
inspection shows that this correlation is driven by a few clusters
with strong Galactic absorption.  As \citet{kaastra2003} pointed
out, in addition to the atomic gas visible at 21~cm these clusters
also have significant X-ray opacity contributions due to dust or
molecules, which naturally explains the observed flux
deficit. When these clusters are excluded, the correlation fades away.

\section{Note added in proof}

Until very recently, the Galactic neutral hydrogen absorption taken
into account when fitting X-ray spectra was extracted from the
\citet[hereafter DL90]{dickey1990} survey.  A better all-sky survey,
the Leiden / Argentine / Bonn (LAB) Galactic \ion{H}{i} Survey, has now become
available \citep{kalberla2005}. Though its use is not yet worldwide,
it has been pointed out that, at least in some cases, the LAB survey gives
notably smaller values than DL90 for the \ion{H}{i} absorption. For example,
in the case of Abell~3112, the LAB value is $1.3 \times
10^{20}$~cm$^{-2}$ while the DL90 value is $2.6 \times
10^{20}$~cm$^{-2}$, and for S\'ersic 159-03 (= AS~1101), the LAB value is
$1.14 \times 10^{20}$~cm$^{-2}$, while the DL90 value is $1.79 \times
10^{20}$~cm$^{-2}$.  Such changes are expected to modify the soft
excess emission derived from spectral fits. Fortunately, for many
other clusters the differences between the \ion{H}{i} absorption measured by
LAB and by DL90 are negligible.

\begin{figure}   
\begin{center}
\psfig{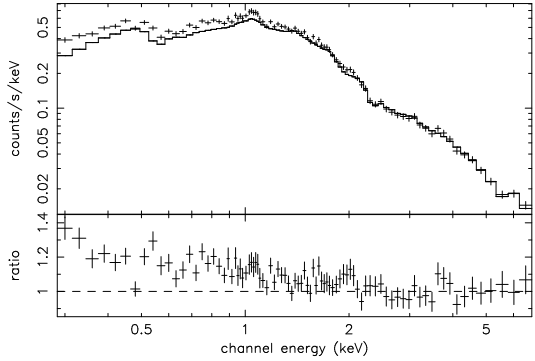}
\caption{XMM-Newton MOS data of Abell~3112 in a $1.5-2.9$ arcmin 
annulus and fit with 
a single temperature MEKAL model in the $2-7$ keV band with the LAB \ion{H}{i}
absorption column, showing a 20~\% soft excess.}
\label{a3112_LAB.fig}
\end{center}  
\end{figure}

For Abell~3112, a fit of the XMM-Newton MOS data in a $1.5-2.9$ arcmin
annulus made by J.~Nevalainen with a single temperature MEKAL model in
the $2-7$ keV band and extrapolated down to 0.3 keV using the LAB \ion{H}{i}
absorption value shows that a 20~\% soft excess remains (see
Fig.~\ref{a3112_LAB.fig}). In order to eliminate the soft excess, the
\ion{H}{i} absorption should take the unrealistically low value of $1.5\
10^{19}$~cm$^{-2}$.

For S\'ersic 159-03, new fits of the joint Suzaku and XMM-Newton data in a $3-8$
arcmin annulus were made by N.~Werner using the new \ion{H}{i} LAB absorption
value. If the absorption is a free parameter in the fit,  N$_{\mathrm H}$ goes
to zero, and no good fit can be obtained with ${\mathrm N}_{\mathrm H} = 1.14\times
10^{20}$~cm$^{-2}$. Adding a thermal component with a temperature of 0.2 keV
improves the fit significantly (from a reduced $\chi ^2$ of 2.8 to 1.5), and the
soft component is found to be present at a significance level of 13.5$\sigma$.
If the fit is made with a multi-temperature model that accounts for the cluster
emission with the absorption left as a free parameter, a 2$\sigma$ upper limit
for N$_{\mathrm H}$ of $1.4\times 10^{19}$~cm$^{-2}$ is obtained; this value is
inconsistent with the new LAB survey and clearly unphysically low. In the same
way, if a power law is fit to the data, even if its total flux is reduced by
about 20~\% it remains significant at a $\sim 10\sigma$ level.  A soft component
is therefore required even for zero absorption, implying that the soft excess is
really strong in this cluster.

Therefore, even if they are somewhat reduced, the soft excesses
observed in Abell~3112 and S\'ersic 159-03 remain when taking into
account the new values of the \ion{H}{i} absorption.

\section{Conclusions}
\label{conclusions}

While in most cases the origin of the soft excess emission is
difficult to prove unambiguously, due to problems with instrument
calibration and unknown background/foreground emission level, the
independent detection of the phenomenon with many instruments gives
confidence in the genuine nature of the phenomenon in a number of
clusters. Both thermal and non-thermal emission mechanisms are
probably at work in producing the soft excess emission in clusters.

\begin{acknowledgements}
The authors thank ISSI (Bern) for support of the team
``Non-virialized X-ray components in clusters of galaxies''.
F.D. acknowledges support from CNES. SRON is supported financially by
NWO, the Netherlands Foundation for Scientific Research. J.N.
acknowledges support from the Academy of Finland. We thank R.~Lieu, 
M.~Bonamente and A.~Fabian for pointing out the
possible modifications of the soft excess emission when using more
recent determinations of the hydrogen absorption along the line of
sight (see note added in proof).
\end{acknowledgements}

\end{document}